\documentclass[conference]{IEEEtran}
\IEEEoverridecommandlockouts
\usepackage{cite}
\usepackage{amsmath,amssymb,amsfonts}
\usepackage{algorithmic}
\usepackage{graphicx}
\usepackage{comment}
\usepackage{textcomp}
\usepackage{xcolor}
\usepackage{siunitx}
\usepackage{amsmath}
\usepackage{algorithm}
\usepackage{url}
\def\BibTeX{{\rm B\kern-.05em{\sc i\kern-.025em b}\kern-.08em
    T\kern-.1667em\lower.7ex\hbox{E}\kern-.125emX}}
    
\title{\LARGE \bf
FLAG: A Framework for \underline{F}PGA-based \underline{L}o\underline{A}d \underline{G}eneration in Profinet 
Communication
}

\author{Ahmad Khaliq$^{1}$, Sangeet Saha$^{1}$, Bina Bhatt$^{1}$, Dongbing Gu$^{1}$, Gareth Howells$^{2}$, 
and Klaus McDonald-Maier$^{1}$
\thanks{This work is funded by INTERREG V 2 SEAS PROJECT INCASE 2S01-049.}
	\thanks{$^{1}$Authors are with Embedded and Intelligent System Lab in School of Computer Science and Electronic Engineering,
   	 University of Essex, Colchester, United Kingdom
		{\tt\small \{ahmad.khaliq,sangeet.saha,bb18131,dgu,kdm\} @essex.ac.uk}}%
	\thanks{$^{2}$Author is with Secure Electronic Systems, University of Kent, United Kingdom
		{\tt\small w.g.j.howells@kent.ac.uk}}
} 
\begin{document}


\maketitle

\begin{abstract}
Like other automated system technologies, PROFINET, a real-time Industrial Ethernet Standard has shown increasing level of integration into the present IT Infrastructure. Such vast use of PROFINET can expose the controllers and I/O devices to operate in  critical failures when traffic goes unexpectedly higher than normal. Rigorous testing of the running devices then becomes essential and therefore, in this paper, we prototype and design an FPGA based load Generating solution called FLAG (FPGA-based LoAd Generator) for PROFINET based traffic at the desired load configurations such as, bits per second, the number and size of the packets with their Ethertypes and MAC addresses. We have employed, a Zynq-7000 FPGA as our implementation platform for the proposed FLAG framework. The system can easily be deployed and accessed via the web or command line interface for successful load generation. Extensive experiments have been conducted to verify the effectiveness of our proposed solution and the results confirm that the proposed framework is capable to generate precise load at Fast/Gigabit line rate with a defined number of packets. 	

\end{abstract}

 \begin{IEEEkeywords}
 \centering
 Profinet, FPGA, NetJury, load generation,
 \end{IEEEkeywords}



\section{Introduction}


The existence of Ethernet in the world of industrial automation and traditional communication systems\cite{dias2017performance} is inevitable. Traditional industrial fieldbuses (such as, Profibus) are relatively slow and due to their limited bandwidth and data sizes, these are now being replaced with real-time industrial Ethernet standard, PROFINET \cite{feld2004profinet}. PROFINET can either be a Component Based Automation, known as PROFINET CBA or PROFINET IO. PROFINET CBA deals with machine to machine communication between distributed automation systems whereas in PROFINET IO, Ethernet based distributed field devices exchange data. This paper is focused on PROFINET IO.

In industrial environment, IO devices and PLC controllers communicate with strict response times \cite{home2014design}. Thus, network bandwidth plays an vital role to fulfill such stringent real-time demands. Performance evaluation of the running automation devices is desirable and requires a scrupulous testing scheme that can help to analyse the performance under various network condition for instance bandwidth \cite{ferrari2011large}. Such scheme utmost requires a framework which can generate network traffic in the form of synthesized packet with real-time attributes and timings. To evaluate the performance of the network and associated devices, this synthesized packets generation procedure can be termed as \textit{load generation} or \textit{net load} \cite{netload}.         

Network load generation is a system which sends synthesized user defined packets into the network while maintaining the desired throughput level. For real-time traffic inspection, common software-based solutions that either run on Windows or Linux have been typically used by the research community. For instance, scapy~\cite{scapy} and pyshark~\cite{pyshark} are python-based software libraries commonly used for packet parsing, live packet capturing and for inspection purposes. Other open-source solutions (such as, netmap~\cite{rizzo2012netmap}) can also be employed and developed for sending user defined traffic while specifying each packet header fields. 

\begin{figure}[t]
    \vspace{-2mm}
	\centering
	\includegraphics[width=1\linewidth, height=1\linewidth,keepaspectratio]{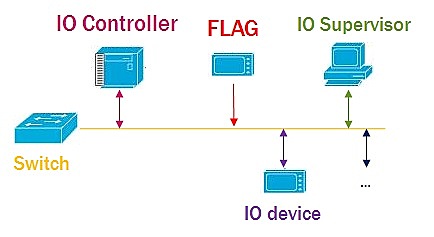}
	\caption{FLAG in Industrial Profinet Communication.}
	\label{figure:flag}
	\vspace{-2mm}
\end{figure}

However, such software-based solutions do not take care of the network available bandwidth and also, due to the involvement of the OS Kernel~\cite{rizzo2012netmap} in allocating memory and handling interrupts for the applications running in the user space, the throughput cannot reach the actual calculated level. In such circumstances, hardware-based solution, such as Field Programmable Gateway Arrays (FPGA) ~\cite{monmasson2007fpga} can be an optimal solution to achieve the real-time measured throughput. Xilinx~\cite{xilinx} being an innovator of hardware based programmable devices, such as FPGAs and SoC, enables the research community to rapidly innovate the ideas with low time cost and faster intelligent computing.

The recent boom of FPGA has encouraged the research community to shift from software to hardware-based solutions. Similar trends have been observed in the industrial automation systems \cite{antolovic2006plc}\cite{monmasson2017fpgas}. In \cite{dias2017performance}, the authors analyzed the performance of PROFINET in motion control application. In \cite{ferrari2011large}, the authors present a case-study of PROFINET IO in industrial automation systems. Furthermore, \cite{durkop2012towards} employed FPGAs for PROFINET software IP cores but utilized the embedded processor (Nios) for the industrial communication. Authors in \cite{flatt2012fpga} for the first time taking advantage of the re-configurable feature of FPGA for PROFINET. ~\cite{nguyen2018fpga} also shown that the FPGA-based implementation of  Media  Access  Control  (MAC)  and  Physical  layer  system (PHY)  for  industrial automation systems is fairly efficient. 

With the advent of recent SoC-based solutions that combines programmable logic units (FPGA) with the Processing system consists of Dual-core ARM Cortex-A9~\cite{wang2011survey} for providing Linux based environments, it becomes relatively easier and efficient to perform time-dependent tasks. Such combination of hardware and software allows the research community to implement the low time cost operations in the Programmable logic unit (PL) while making use of PS for triggering those operations with the support of programming language including Python, C/C++, VHDL etc. Taking precedence of the re-configurable nature coupled with hardware and software subsystems of Xilinx NetJury \cite{netJuryXilinx}, we have employed it as our implementation platform and proposed a hardware-based solution ``FLAG'' for PROFINET based traffic generation, as shown in Figure \ref{figure:flag}. Several experimentation shows its effectiveness in generating desired amount of the load with compact time limitations in PROFINET communication. The main contribution of this paper can be stated as:
\begin{itemize}
    \item Proposed a framework, ``FLAG'' for generating network traffic with desired throughput/load in PROFINET communication.
    \item Integrated the framework with re-configurable FPGA-based NetJury device for user defined load generation on the fly.
    \item Precise experimental validation utilizing Hilscher netANALYZER and tektronix DPO4054B oscilloscope-based analysis reveals the efficiency of our proposed framework in real-time industrial communication.
\end{itemize}

The remainder of the paper is organized as follow. Section II illustrates the brief overview of the FLAG framework. Section III presents an explanation of the implementation platform followed by the experimentation validation carried out in Section IV. The conclusion is presented in Section V.

\section{FLAG Framework}
This section will discuss the functionality of the proposed FLAG framework. In-short, our proposed architecture sends user defined packets into the network while achieving a desired throughput, also named as network load. We will first discuss the fundamental structure of a packet, followed by the discussion on the load generation scheme. 

\subsection{Fundamental of a Network Packet}
A network packet is a formatted unit of data carried by a packet-switched network \cite{akerberg2009exploring}. A packet is generally composed of control information and user data, often referred as header and payload. Delivering the payload depends upon the header information which includes source and destination network or MAC address along with other sequencing information. A basic fields of a packet is shown in Table \ref{table:packets}.    

\renewcommand{\arraystretch}{1}
\renewcommand{\tabcolsep}{2pt}
\begin{table}[h]
\centering
\vspace{-2mm}
\caption{Parameters of a packet}
\label{table:packets}
\begin{tabular}{|l|l|l|}
\hline
\textbf{Field}      & \textbf{Length} & \textbf{Description}                                                           \\ \hline
Preamble $p$            & 7 bytes         & Synchronizes comminication                                                     \\ \hline
Start of Frame $d$      & 1 byte          & Signals the start of a valid frame                                             \\ \hline
MAC Destination $M_d$     & 6 bytes         & Destination MAC address                                                        \\ \hline
MAC Source $M_s$         & 6 bytes         & Source MAC address                                                             \\ \hline
802.1Q tag  $v$        & 4 bytes         & Optional VLAN tag                                                              \\ \hline
Ethertype or length $E$ & 2 bytes         & Payload type or frame size                                                     \\ \hline
Payload $P$            & 42-1500 bytes   & Data payload                                                                   \\ \hline
CRC $f$                & 4 bytes         & Frame error check                                                              \\ \hline
Interframe Gap $i$      & 12 bytes        & \begin{tabular}[c]{@{}l@{}}Required idle Period between \\ frames\end{tabular} \\ \hline
\end{tabular}
\end{table}

Packet fields including Payload $P$ with other fixed sized MAC addresses $M \in \{M_d,M_s\}$, Ethertype $E$, preamble $p$, frame start/delimiter $d$, frame check/crc $f$ and optional $v$ vLAN tag collectively represent a $F$ frame with size $S$ shown in equation~(\ref{eq:0}). 
\vspace{-4mm}

\begin{equation}
\vspace{-1mm}
    S = \{p, d, M, v, E, P, f\}
    \label{eq:0}
\end{equation}

In this paper, by varying payload $P$, the overall untagged vLAN frame size $S \in \{72,136,268,524,1036,1526\}$. However, if the frame is vLAN tagged, an extra $4$ byte will be further added in $S$. 

\subsection{Load Generation strategy}
Considering modern industrial automation devices, our proposed FLAG should be able to maintain the desired throughput, while allowing normal operation of the associated devices. Therefore, the desired throughput or network load should be less than the minimum bandwidth required by the network and devices. To maintain a $L$ \% of network load, the FLAG will send $F$ number of frames. To make sure industrial controllers and IO devices keep operating normally, we have devised a strategy to send frames with gap $I_L$, called $Load \ Gap$. This $I_L$ can be calculated as:
\vspace{-2mm}
\begin{equation}
    I_L = 12 + (12 + S) \times (\frac{1-L}{L})
    \label{eq:loadGap}
\end{equation}

All $F$ frames will have $I_L$ gap in bytes. In equation~\ref{eq:loadGap}, $12$ represents the minimum gap, $L$ denotes the desired load percentage where $S$ be the frame size computed using (\ref{eq:0}). At $L=100 \%$, $I_L$ will be $12$ and all the network bandwidth will be going to be consumed by FLAG, therefore, it is important to set the $L$.  

In real-time PROFINET communication, generating $L$ load into the network with large number of $F$ frames might take an infinite time. Therefore, it will be much more convenient if the user can be informed about the amount of time $T$ within which the packets/load will be transmitted. This $T$ time can be calculated using equation~(\ref{eq:2}).

\begin{equation}
   T = \frac {(S+12) \times 8 \times F} {L \times R} , \ R \in \{100,1000\}
   \label{eq:2}
\end{equation}

\begin{figure}[htbp]
	\centering
    \includegraphics[width=1\linewidth, height=0.5\linewidth,keepaspectratio]{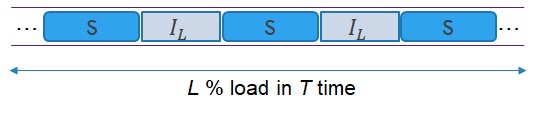}
\caption{$S$ sized $F$ frames with $I_L$ load gap for $L$ \% load in $T$ time.}
\label{figure:pkt_load}
\end{figure}

In equation~\ref{eq:2}, $R$ denotes the total network bandwidth which can either be $100$ for Fast Ethernet or $1000$ for Gigabit Ethernet. A minimum interframe gap of $12$ is added into the $S$ frame size and converted into bits by multiplying it with $8$. Thus, $T$ will be the actual time in seconds for FLAG to generate $L$ \% load by sending $F$ number of frames into the industrial communication illustrated in Figure~\ref{figure:pkt_load}.

\section{FLAG on Zynq-7000 FPGA series}
In this section, we will first go deep into the FPGA-based ``NetJury'' architecture employed for the implementation of the proposed FLAG. Later, we will discuss how the proposed FLAG framework mapped inside the NetJury. 

\subsection{NetJury background and its functionality}
\begin{figure}[htbp]
	\centering
    \includegraphics[width=1\linewidth, height=1\linewidth,keepaspectratio]{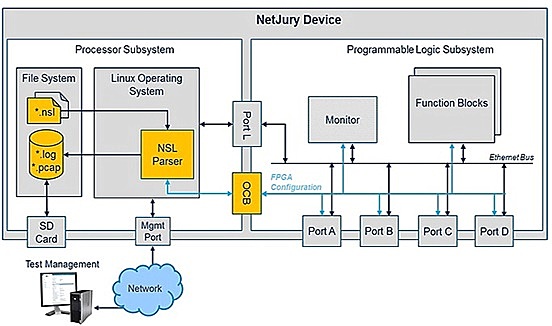}
\caption{Architectural view of the NetJury \cite{netJuryXilinx}.}
\label{figure:NetJury}
\end{figure}

For the execution of the proposed FLAG framework, we have employed Xilinx Zynq FPGA based NetJury device \cite{netJuryXilinx}  by Netmodule \cite{netmodule} in collaboration with Xilinx \cite{netJuryXilinx}. It is based on Zynq All Programmable SoC integrated with a Dual-core ARM Cortex-A9 processing subsystem~\cite{wang2011survey}. 
The PL (FPGA) is designed to process real-time data and the Linux-based processing subsystem is used to execute the desired functionality in the PL unit. To program the PL subsystem, NetJury supports a proprietary scripting language, named as NetJury Scripting Language (NSL). A block diagram of the NetJury is shown in Figure \ref{figure:NetJury}.    

\vspace{-2mm}

In Figure \ref{figure:NetJury}, the programming logic (PL) unit is on the right side and the left side is the Linux based dual-core ARM processor. The processor Subsystem (PS) can be accessed via Putty~\cite{pearson1974putty} or Winscp~\cite{winscpscp} utilizing the console port or through the Network via management port. Four Ethernet ports running at Fast/Gigabit bandwidth are provided for real-time Ethernet testing. Other components of the PL subsystem contain a Monitor module and hidden blocks containing registers. For each Ethernet port, packet generation with manipulation up-to the application layer can be performed via NSL scipt. The Port L (eth1) being a bridge port, communicate information between both the subsystems. With user defined NSL script(s), the PS can trigger the desired functionality in PL subsystem. Standard high-level programming languages, including Python, can be employed for the generation and execution of the NSL scripts. Upon execution of any test initiate by the user (if defined in the NSL) a log file and pcap file~\cite{deri2004improving} generated in the PS block. The pcap file contains the actual packets being send/received on any of the FPGA ports. Within the Linux OS, pre-installed standard packet processing tools including scapy, iperf \cite{ferrari2008testing} and communication protocol stacks like IEC 61850\cite{mackiewicz2006overview} and PROFINET are provided for packets analysis and generation.   

\subsubsection{Netjury Scripting Language (NSL)}
\begin{figure}[h]
	\centering
	\vspace{-2mm}
    \includegraphics[width=1\linewidth, height=1\linewidth,keepaspectratio]{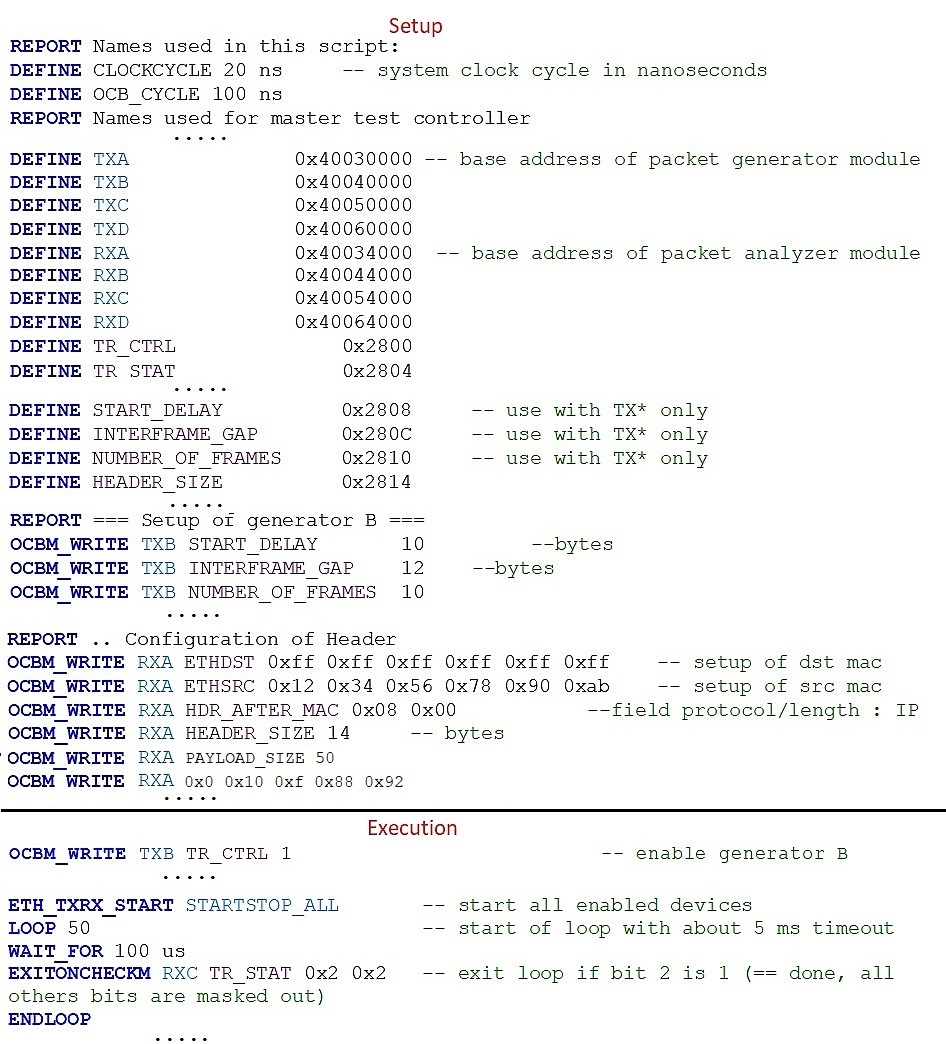}
\caption{Sample NSL script with Setup and Execution phase.}
\label{figure:setup}
\end{figure}

Typical NSL commands are shown in Table \ref{table:nsl_commands}. The NSL script is generally partitioned into three main phases (please see Figure \ref{figure:setup}):
\begin{itemize}
\item Setup Phase: Initialization of the Ethernet frame generators and analyser.
\item Execution Phase: Ethernet frames are generated, and the analyser is enabled - the components run with high determinism and fulfill real-time requirements.
\item Reporting Phase: The results of the execution phase are collected from each frame analyser instance and written to the log file in PS.
\end{itemize}
\vspace{-3mm}
\begin{table}[ht]
\centering
\caption{Basic NSL commands\cite{netJuryXilinx}}
\label{table:nsl_commands}
\begin{tabular}{|l|l|}
\hline
\textbf{Command} & \textbf{Meaning}                        \\ \hline
REPORT           & Description of the script and test case \\ \hline
DEFINE           & Definition of constants and registers   \\ \hline
REF              & Test case reference                     \\ \hline
OCBM\_WRITE      & Setup of Generators and Analysers       \\ \hline
ETH\_TXRX\_START & Start of execution phase                \\ \hline
LOOP             & Start of loop                           \\ \hline
WAIT\_FOR        & Delay in time or cycles                 \\ \hline
EXITONCHECK      & Conditional exit                       \\ \hline
END LOOP         & End of loop                             \\ \hline
OCBM \_CHECK     & Analyse results                         \\ \hline
ETH\_TXRX\_STOP  & End of Execution phase                  \\ \hline
\end{tabular}
\end{table}

Using the commands in the Table \ref{table:nsl_commands}, parameters including the addresses of the registers, clockcycle, etc should be written in the NSL script. A pythonic NetJury library for NSL generation and execution is written for fast development and testing purpose.

\subsubsection{Modes of the NetJury}

There are three main modes of the NetJury;  1) Transparent mode, a default mode in which the traffic between all FPGA Ethernet ports can be accessed at Port L with no FPGA programming access. 2) Switching mode creates a bridge between Port A and Port B of the NetJury i.e. traffic received at Port A send out of Port B and vice-versa. 3) Scripting mode allows the user to program the PL unit for traffic generation and manipulation (if specified). 

\begin{figure}[htbp]
	\centering
	\vspace{-2mm}
    \includegraphics[width=1\linewidth, height=1\linewidth,keepaspectratio]{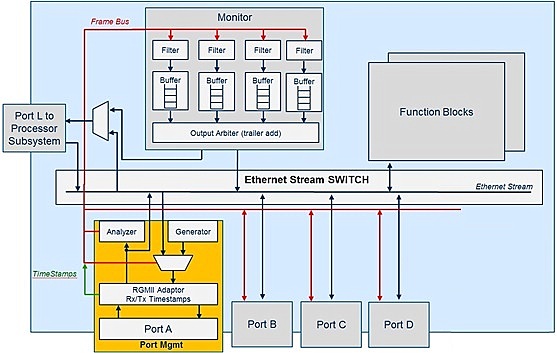}
\caption{Structure of PL subsystem \cite{netJuryXilinx}.}
\label{figure:PL}
\end{figure}

\subsubsection{Packet Generator and Analyser}
Each Ethernet port consists of a packet generator that triggers the packet sending process and an analyser which allows the packet manipulation. Both the generator and analyser are controlled via NSL (with an NSL parser to decode it) which setup the PL registers through an On-Chip Bus (OCB). A detailed structural view of the Programming Subsystem is shown in Figure \ref{figure:PL}. A frame received at any of the Ethernet port is timestamped, computed at the arrival of start frame delimiter in the PHY clock domain with the compensation of the buffer latency. Its precision depends upon the dedicated clock of $8ns$ (system clock is $125MHz$). Traffic received at any of the Ethernet port can take multiple paths; under the switching mode, the Ethernet Stream divert it to other ports. 
Under the scripting mode, if enabled, the analyser inspects each packet up-to application layer, matches the user defined patterns, trigger the frames counters and append the frame with some event specific code tag (if tagging is enabled). Finally, it generates the statistics of the received traffic for automatic test evaluation and report generation. The Ethernet frame analyser attributes with their description are shown in Table \ref{table:frame_analyser_attrib}.

\renewcommand{\arraystretch}{1.1}
\renewcommand{\tabcolsep}{1pt}

\begin{table}[h]
\vspace{-3mm}
\centering
\caption{Frame analyser attributes\cite{netJuryXilinx}}
\label{table:frame_analyser_attrib}
\begin{tabular}{|l|l|}
\hline
\textbf{Attribute}    & \textbf{Description}                                                                                                                    \\ \hline
TRANSMITTER\_CONTROL  & Disable or Enable                                                                                                                        \\ \hline
TRASMITTER\_STATE     & Disable or Receiving | Hold                                                                                                              \\ \hline
FRAMES\_EXP           & \begin{tabular}[c]{@{}l@{}}Total number of frames to be received until \\ the Analyser will be automatically stopped\end{tabular}       \\ \hline
FRAMES\_EXP\_OK       & \begin{tabular}[c]{@{}l@{}}Number of expected frames to be received \\ until the Analyser will be automatically \\ stopped\end{tabular} \\ \hline
NUMBER\_OF\_RECV\_OK  & Number of matching frames received                                                                                                      \\ \hline
NUMBER\_OF\_RECV\_NOK & Number of frames received with mismatch                                                                                                 \\ \hline
ERROR\_CODE           & \begin{tabular}[c]{@{}l@{}}Error code to be inserted into trailer frame\\  is forwarded\end{tabular}                                    \\ \hline
\end{tabular}
\end{table}

Any frame which goes out of the NetJury's Ethernet port either comes from the generator under scripting mode or Ethernet stream path under switching mode. 
Under the scripting mode, the user needs to define the packet header parameters inside the NSL script. Features like calling specific registers to dynamically modify next frame data or loops for deterministic frame generation at wire-rate, all should be predefined inside the NSL. The Ethernet frame generator attributes are shown in Table \ref{table:frame_generator_attrib}.

\begin{table}[ht]
\vspace{-5mm}
\centering
\caption{Frame generator attributes\cite{netJuryXilinx}}
\label{table:frame_generator_attrib}
\begin{tabular}{|l|l|}
\hline
\textbf{Attribute}   & \textbf{Description}                             \\ \hline
TRANSMITTER\_CONTROL & Disable or Enable                                 \\ \hline
TRASMITTER\_STATE    & Disable or Transmitting or Done                    \\ \hline
INTERFRAME\_GAP      & Gap between transmitted packet                   \\ \hline
START\_DELAY         & Delay from reference point                       \\ \hline
NUMBER\_OF\_FRAMES   & \begin{tabular}[c]{@{}l@{}}Number of frames transmitted \\ with INTERFAME\_GAP \\\end{tabular} \\ \hline
\end{tabular}
\end{table}

\subsubsection{Monitor}
A monitor module is designed for accessing the traffic at port L either send by the generators or received by the analysers. There are two main components inside the monitor; 1) Filter that can remove undesired frames by specifying the source and destination MAC addresses and Ethertypes. When receiving traffic from multiple ports, at Port L, no frame usually has port information. In filters, enabling the port tag field feature, the user can append the packets with port information, 2) Buffer is the another component within which the incoming frames are stored. The monitor module is currently restricted to store 4 frames only, thus, we cannot expect to receive 100\% traffic for analysis purpose.

\subsection{FLAG on NetJury}
To employ the NetJury device for FLAG framework, we go deep into the proprietary NSL script and develop it to send user defined number of frames at desired load configurations. A sample image of the NSL script is shown in Figure \ref{figure:setup}. Inside the setup block, the user needs to define the addresses of the registers including generators and analysers of the ports. Such information will be same for all the NSL scripts. Following these, the packet header fields such as destination and source MAC addresses, Ethertype and Payload size are then defined. The execution phase includes triggering the registers to start and stop sending packets. For instance, setting the \textit{TR\_CTRL} to $1$ amkes the  specific port ready to send the frames. \textit{ETH\_TXRX\_START} start sending the frames with \textit{LOOP} and \textit{WAIT\_FOR} control the delay. \textit{EXITONCHECKM} is for conditional exit but \textit{ETH\_TXRX\_STOP} can also be used to stop sending frames otherwise. Since, we are interested in generating traffic only with no intention of analysing the response packets, therefore, we have not employed the analyser module (monitor) for load generation. Therefore, the user requires to provide the following fields to NetJury to be able to generate the desired load.

\begin{itemize}
	\item \textit{Source MAC address}
	\item \textit{Destination MAC address}
	\item \textit{Ethertype}
	\item \textit{Payload size}
	\item \textit{Number of frames}
	\item \textit{Interframe gap}
\end{itemize}

\section{Experimentation Validation}
This section elaborates on the performance efficiency of our  \textit{FLAG} framework in real-time traffic generation. In the PS side, packets are synthesized by using pythonic modules including Scapy and converted into NSL syntax (see Figure \ref{figure:setup}) for load generation. Furthermore, Hilscher netANALYZER~\cite{HilscherAnalyzer} and a tektronix DPO4054B Oscilloscope~\cite{tektro} are employed for accurate measurements.  

\subsubsection{Command Line Interface (CLI) for Load generation}
\begin{figure}[h]
	\centering
	\includegraphics[width=1\linewidth, height=1\linewidth,keepaspectratio]{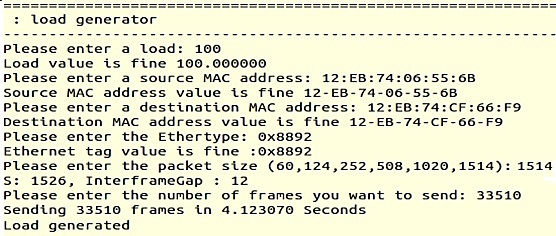}
	\caption{100\% load with 33510 frames}
	\label{figure:100_load_non_hmi}
\end{figure}

\begin{figure}[h]
	\centering
	\includegraphics[width=1\linewidth, height=1\linewidth,keepaspectratio]{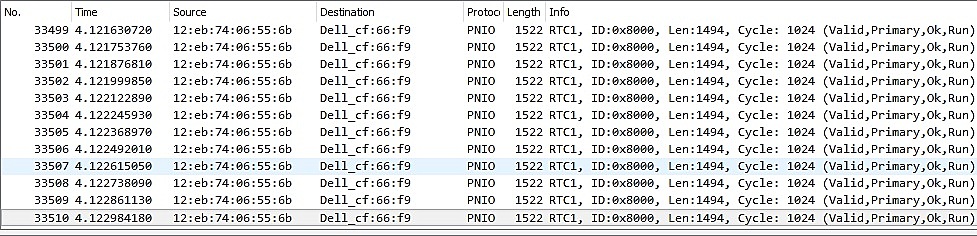}
	\caption{Hilscher netANALYZER measurement for 100\% load with 33510 frames.}
	\label{figure:100_load_wiresh}
\end{figure}

To generate packets while maintaining load at desired \%, we designed a console-based interface which can either run on Linux terminal or Putty running on Windows. The user needs to specify the percentage of load, packet headers' fields, packet size and number of frames to be send. In Figure~\ref{figure:100_load_non_hmi}, the user specified 100\% load with packet size $1514$ and $33510$ frames. Therefore, it makes $S=1526$, $I_L=12$ and $T=4.12370s$, also shown in Figure~\ref{figure:100_load_non_hmi}. Once, the FLAG generates the load, Figure~\ref{figure:100_load_wiresh} illustrates the number of total packets captured using Hilscher netANALYZER. It is quite evident that the last frame $33510^{th}$ is captured at $4.1229s$ which is equal to the expected $T$. 

\begin{figure}[h]
	\centering
	\includegraphics[width=1\linewidth, height=1.3\linewidth,keepaspectratio]{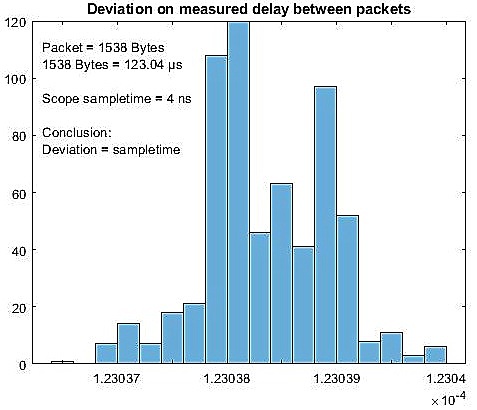}
	\caption{Oscilloscope measurement for 100\% load with 33510 frames.}
	\label{figure:100_load_oscil}
\end{figure}

Furthermore, MATLAB decoding of the packets captured using Oscilloscope is performed. Figure~\ref{figure:100_load_oscil} illustrates the deviation delay measured between the captured packets. With overall packet size of $1538$ bytes, time per frame comes around $T_f=123.04 \mu s$ using equation~(\ref{eq:fr_end}). With oscilloscope, $F=624$ packets are captured during a 80ms measurement with start time, $t_0: 0 s$ and
end time, $t_1: 0.076653 s$, the overall load can be determined using equation~(\ref{eq:load}), comes around 100\% load.  
\vspace{-1mm}
\begin{equation}
    T_f = \frac{(S+12) \times 8} {R}
   \label{eq:fr_end}
\end{equation}

\begin{equation}
Load = \frac {(S+12) \times 8 \times F} {t1-t0}
\label{eq:load}
\end{equation}
\vspace{-2mm}

\vspace{-2mm}

\begin{figure}[!h]
	\vspace{-2mm}
	\centering
	\includegraphics[width=1\linewidth, height=0.9\linewidth,keepaspectratio]{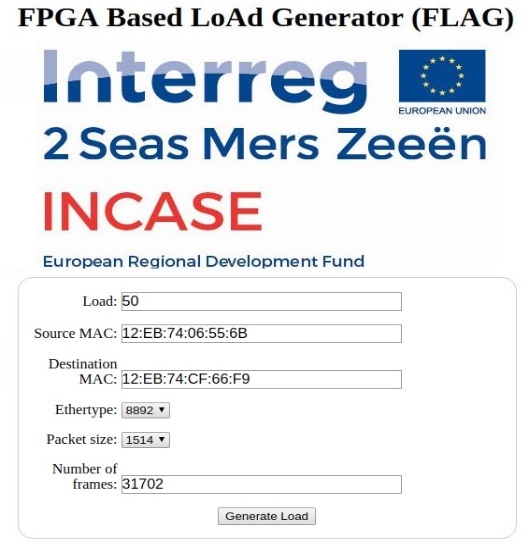}
	\caption{50\% load with 31702 frames.}
	\label{figure:50P_load_hmi}
\end{figure}

\begin{figure}[!h]
    \vspace{-2mm}
	\centering
	\includegraphics[width=1\linewidth, height=1.3\linewidth,keepaspectratio]{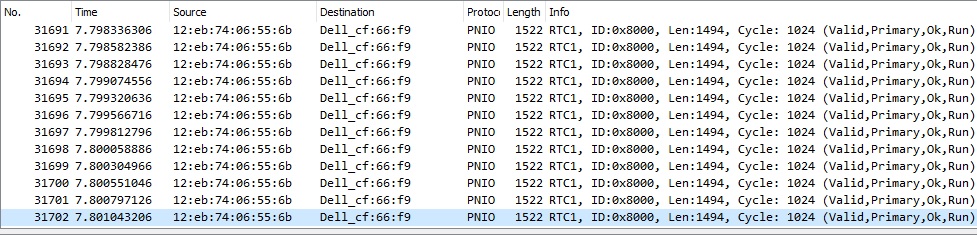}
	\caption{Hilscher netANALYZER measurement for 50\% load with 31702 frames.}
	\label{figure:50_load_wiresh}
\end{figure}

\subsubsection{Human Machine Interface (HMI) for Load generation}
For ease of access, we designed a Human Machine Interface (HMI) for load generation shown, in Figure \ref{figure:50P_load_hmi}, where the user can enter the desired load configurations. In this experiment shown, the user specified 50\% load with $1514$ packet and $31702$ frames. Again, it makes $S=1526$, $I_L=1550$ and $T=7.8012s$. Figure \ref{figure:50_load_wiresh} confirms that the experimental time duration is equal to the calculated $T$. The last $31702^{th}$ frame captured at $7.801s$, thus, $t_1-t_0=7.801s$ and $F=31702$, the load from equation (\ref{eq:load}) approximately comes around $50\%$ load.

From the above test scenarios, it is evident that FLAG is efficient in generating real-time Profinet traffic. The load \% can vary between $0$ and $100$. Also, vLAN tagged packets can also be generated with the proposed framework by changing the \textit{HDR\_AFTER\_MAC} and \textit{PAYLOAD\_SIZE}. Our proposed FLAG framework can furthermore be used for generating load other than PROFINET by changing the Ethernet type / \textit{HDR\_AFTER\_MAC}. 

\section{Conclusion}
This paper presents a prototype and the design a hardware based solution ``FLAG" for thorough testing of Industrial devices in real-time PROFINET communication. As an implementation platform, we integrate ``NetJury" Zynq-7000 FPGA which is Zynq All Programmable SoC and proposed an architecture that can generates load by sending packets into the industrial network. For precise and accurate measurements, we have used Hilscher netANALYZER and tektronix DPO4054B Oscilloscope. Several experiments have shown that the proposed FLAG framework is capable to generate a constant load with defined number of packets. FLAG can be accessed either via terminal of Linux or Windows. Alongside, an HMI interface is also developed for ease of use.

\section*{ACKNOWLEDGMENT}

We gratefully acknowledge the support of Frederic Depuydt, Philippe Saey as well as the INCASE project partners for sharing knowledge. 
\newpage
\bibliographystyle{ieeetr}
\bibliography{root}

\begin{thebibliography}{10}

\bibitem{dias2017performance}
A.~L. Dias, G.~S. Sestito, and D.~Brandao, ``Performance analysis of profibus
  dp and profinet in a motion control application,'' {\em Journal of Control,
  Automation and Electrical Systems}, vol.~28, no.~1, pp.~86--93, 2017.

\bibitem{feld2004profinet}
J.~Feld, ``Profinet-scalable factory communication for all applications,'' in
  {\em IEEE International Workshop on Factory Communication Systems, 2004.
  Proceedings.}, pp.~33--38, IEEE, 2004.

\bibitem{home2014design}
S.~H{\"o}me, S.~Palis, and C.~Diedrich, ``Design of communication systems for
  networked control system running on profinet,'' in {\em 2014 10th IEEE
  Workshop on Factory Communication Systems (WFCS 2014)}, pp.~1--8, IEEE, 2014.

\bibitem{ferrari2011large}
P.~Ferrari, A.~Flammini, F.~Venturini, and A.~Augelli, ``Large profinet io rt
  networks for factory automation: A case study,'' in {\em ETFA2011}, pp.~1--4,
  IEEE, 2011.

\bibitem{netload}
``Netload.''
  \url{https://de.profibus.com/downloads/profinet-security-level-1-netload/}.

\bibitem{scapy}
``Scapy.'' \url{https://scapy.readthedocs.io/en/latest/}.

\bibitem{pyshark}
``pyshark.'' \url{https://pypi.org/project/pyshark/}.

\bibitem{rizzo2012netmap}
L.~Rizzo, ``Netmap: a novel framework for fast packet i/o,'' in {\em 21st
  USENIX Security Symposium (USENIX Security 12)}, pp.~101--112, 2012.

\bibitem{monmasson2007fpga}
E.~Monmasson and M.~N. Cirstea, ``Fpga design methodology for industrial
  control systems—a review,'' {\em IEEE transactions on industrial
  electronics}, vol.~54, no.~4, pp.~1824--1842, 2007.

\bibitem{xilinx}
``Xilinx.'' \url{https://www.xilinx.com/}.

\bibitem{antolovic2006plc}
M.~Antolovic, K.~Acton, N.~Kalappa, S.~Mantri, J.~Parrott, J.~Luntz, J.~Moyne,
  and D.~Tilbury, ``Plc communication using profinet: experimental results and
  analysis,'' in {\em 2006 IEEE Conference on Emerging Technologies and Factory
  Automation}, pp.~1--4, IEEE, 2006.

\bibitem{monmasson2017fpgas}
E.~Monmasson, ``Fpgas: Fundamentals, advanced features, and applications in
  industrial electronics [book news],'' {\em IEEE Industrial Electronics
  Magazine}, vol.~11, no.~2, pp.~73--74, 2017.

\bibitem{durkop2012towards}
L.~D{\"u}rkop, H.~Trsek, J.~Jasperneite, and L.~Wisniewski, ``Towards
  autoconfiguration of industrial automation systems: A case study using
  profinet io,'' in {\em Proceedings of 2012 IEEE 17th International Conference
  on Emerging Technologies \& Factory Automation (ETFA 2012)}, pp.~1--8, IEEE,
  2012.

\bibitem{flatt2012fpga}
H.~Flatt, S.~Schriegel, T.~Neugarth, and J.~Jasperneite, ``An fpga based hsr
  architecture for seamless profinet redundancy,'' in {\em 2012 9th IEEE
  International Workshop on Factory Communication Systems}, pp.~137--140, IEEE,
  2012.

\bibitem{nguyen2018fpga}
T.~T.~T. Nguyen, Y.~Nagao, T.~Uwai, N.~Sutisna, M.~Kurosaki, H.~Ochi, and
  B.~Sai, ``Fpga implementation of wireless lan system for factory
  automation,'' in {\em 2018 International Conference on Advanced Technologies
  for Communications (ATC)}, pp.~78--83, IEEE, 2018.

\bibitem{wang2011survey}
W.~Wang and T.~Dey, ``A survey on arm cortex a processors,'' {\em Retrieved
  March}, 2011.

\bibitem{netJuryXilinx}
``Netjury website.''
  \url{https://www.xilinx.com/products/boards-and-kits/1-4z9mkv.html}.

\bibitem{akerberg2009exploring}
J.~Akerberg and M.~Bjorkman, ``Exploring security in profinet io,'' in {\em
  2009 33rd Annual IEEE International Computer Software and Applications
  Conference}, vol.~1, pp.~406--412, IEEE, 2009.

\bibitem{netmodule}
``Netmodule.'' \url{http://www.netmodule.com/netmodule-home.html}.

\bibitem{pearson1974putty}
R.~Pearson, ``Putty application tool,'' July~2 1974.
\newblock US Patent 3,821,828.

\bibitem{winscpscp}
S.~WinSCP-Free, ``Scp and ftp client for windows.''

\bibitem{deri2004improving}
L.~Deri {\em et~al.}, ``Improving passive packet capture: Beyond device
  polling,'' in {\em Proceedings of SANE}, vol.~2004, pp.~85--93, Amsterdam,
  Netherlands, 2004.

\bibitem{ferrari2008testing}
P.~Ferrari, A.~Flammini, D.~Marioli, S.~Rinaldi, and A.~Taroni, ``Testing
  coexistence of different rte protocols in the same network,'' in {\em 2008
  IEEE International Workshop on Factory Communication Systems}, pp.~179--187,
  IEEE, 2008.

\bibitem{mackiewicz2006overview}
R.~E. Mackiewicz, ``Overview of iec 61850 and benefits,'' in {\em 2006 IEEE PES
  Power Systems Conference and Exposition}, pp.~623--630, IEEE, 2006.

\bibitem{HilscherAnalyzer}
``Hilscher netanalyzer.'' \url{www.de.hilscher.com}.

\bibitem{tektro}
``tektronix dpo4054b.'' \url{http://www.tektronix.com }.

\end{thebibliography}
\end{document}